\documentclass[fleqn,twoside]{article}
\usepackage[headings]{espcrc2}

\readRCS
$Id: espcrc2.tex,v 1.2 2004/02/24 11:22:11 spepping Exp $
\usepackage{graphicx}
\usepackage[figuresright]{rotating}

\newcommand{\AmS}{{\protect\the\textfont2
  A\kern-.1667em\lower.5ex\hbox{M}\kern-.125emS}}

\def\MgB2{MgB$_{2}$}
\def\cm-1{cm$^{-1}$\,}
\def\cmT-1{cm$^{-1}$/T\,}
\def\E2g{$E_{2g}$}
\def\A1g{$A_{1g}$}
\def\2DS{$2\Delta_{S}^{E}$}

\def\2DA{$2\Delta^{A}$}
\def\D0{$2\Delta_{0}$}

\title{Multi-Gap Superconductivity in \MgB2: Magneto-Raman Spectroscopy}

\author{G.~Blumberg\address[BL]{Bell Laboratories, Lucent 
Technologies, Murray Hill, NJ 07974, USA}\thanks{Corresponding 
author. E-mail: girsh@bell-labs.com}, 
A. Mialitsin\addressmark\thanks{Department of Physics and Astronomy, 
Rutgers University, Piscataway, NJ 08854, USA},
	 B. S. Dennis\addressmark,
       N.~D.~Zhigadlo\address[ETH]{Solid State Physics Laboratory, ETH, 
       CH-8093 Z\"urich, Switzerland} 
        and
        J.~Karpinski\addressmark}
       
\runtitle{Multi-Gap Superconductivity in \MgB2: Magneto-Raman Spectroscopy}
\runauthor{G.~Blumberg {\it et. al.}}

\begin{document}

\begin{abstract}
Electronic Raman scattering studies on \MgB2 single crystals as a 
function of excitation
and polarization have revealed three distinct superconducting
features: a clean gap below 37~\cm-1 and
two coherence peaks at 109~\cm-1 and 78~\cm-1 which we identify as 
the superconducting gaps in $\pi$- and $\sigma$-bands and as the 
Leggett's collective mode arising from the 
fluctuation in the relative phase between two superconducting 
condensates residing on corresponding bands. 
The temperature and field dependencies of the superconducting features
have been established. 
A phononic Raman scattering study of the \E2g boron stretching mode 
anharmonicity 
and of superconductivity induced self-energy effects is presented. 
We show that anharmonic two phonon decay is mainly responsible for the
unusually large linewidth of the \E2g mode. 
We observe $\sim 2.5\%$
hardening of the \E2g phonon frequency upon cooling into the 
superconducting state and estimate the electron-phonon coupling
strength associated with this renormalization.
\vspace{1pc}
\end{abstract}

\maketitle

\section{INTRODUCTION}

The multi-gap nature of superconductivity in \MgB2 was predicted 
theoretically \cite{Liu} and has been 
experimentally established by a number of spectroscopies. 
A double-gap structure in the quasi-particle energy spectra was 
determined from tunneling spectroscopy \cite{Iavarone,Sza30}. 
The two gaps have been assigned to distinctive quasi-two-dimensional 
$\sigma$-bonding states of the boron $p_{x,y}$ orbitals and 
three-dimensional $\pi$-states of the boron $p_z$ orbitals Fermi 
surface (FS) sheets by means of ARPES~\cite{Tsuda,Souma}:  
$\Delta_{\sigma} =  5.5 - 6.5$ and $\Delta_{\pi} = 1.5 - 
2.2$~meV. 
Scanning tunneling microscopy (STM) has provided a reliable fit for 
the smaller gap, $\Delta_{\pi} = 2.2\,{\rm meV}$ \cite{Esklidsen}. 
This value manifests in the absorption threshold
energy at 31~\cm-1 obtained from
magneto-optical far-IR studies~\cite{Perucchi}. 
The nominal upper critical field $H^{\pi}_{c2}$ deduced from the
coherence length $\xi_{\pi} = 49.6\,{\rm nm}$ by vortex imaging is
$H^{\pi}_{c2} \approx 0.13\,{\rm T}$~\cite{Esklidsen} which is
much smaller than the critical field $H^{opt}_{c2} \approx
5\,{\rm T}$ found by magneto-optical measurements~\cite{Perucchi}.

Electronic Raman studies on \MgB2 have explored the
superconducting (SC) energy gap and changes in phonon lineshapes, starting 
with the work of \cite{Chen,Goncharov} and followed thereafter by
\cite{Quilty,TajimaCax}. 
The dependence of the Raman response on scattering geometry allowed 
an observation of the pairing gap on
the two-dimensional $\sigma$ bands and the 3D $\pi$ bands. 
By orienting the light polarizations along the c-axis of \MgB2
(perpendicular to the hexagonal planes) the weakly dispersing 
$\sigma$ bands cannot be probed and thus only the $\pi$ bands
are projected out, giving an observed threshold at 
$2\Delta_{\pi}=29$~\cm-1 \cite{TajimaCax}. 
The larger $2\Delta_{\sigma}$ gap has been demonstrated by Raman 
experiments as a SC coherence peak at 105~\cm-1\,\cite{Quilty}.  

For multi-band superconductors collective modes 
associated with fluctuations of the relative phase and amplitudes of 
coupled condensates \cite{Leggett,Griffin,Marel} as well as distinctive 
self energy effects associated with intra- and inter-band 
interactions \cite{Liu,Mazin} were expected. 
It has been suggested from STM vortex imaging that the 
superconductivity in the $\pi$-band is induced by 
superconductivity in the $\sigma$-band \cite{Esklidsen}, however, the 
coupling mechanism remained unclear. 
Previous phononic Raman spectroscopy has identified a broad
$\Gamma$-point phonon centered around $620-640$~\cm-1   
\cite{Chen,Goncharov,Quilty} consistent with the calculated frequency 
of the anharmonic \E2g boron stretching 
mode~\cite{Mazin,Yildirim}. 
The phononic dispersion has been studied by  
inelastic x-ray scattering \cite{Shukla,Baron}.  
However, the expected self energy effects \cite{Liu,Mazin} have not 
been demonstrated.   

\subsection{Experimental}

Polarized Raman scattering can probe excitations around the 
Brillouin zone (BZ) center that belong to different symmetry 
representations within the space group of the crystal structure. 
The point group associated with ${\rm MgB_2}$ is $D_{6h}$.
We denote by
$(\textbf{e}_{in} \textbf{e}_{out})$ a configuration in which the
incoming/outgoing photons are polarized along the
$\textbf{e}_{in}$/$\textbf{e}_{out}$ directions. The vertical
($V$) or horizontal ($H$) directions were chosen
perpendicular or parallel to the crystallographic $a$-axis. The
''right-right'' ($RR$) and ''right-left'' ($RL$)
notations refer to circular polarizations:
$\textbf{e}_{in} = (H - i V) / \sqrt{2}$, with
$\textbf{e}_{out} = \textbf{e}_{in}$ for the $RR$ and
$\textbf{e}_{out} = \textbf{e}_{in}^{*}$ for the $RL$ geometry. 
For the $D_{6h}$ point group the $RR$ and $H\!H$
polarizations select correspondingly $A_{1g}$ and $A_{1g}$ +
$E_{2g}$ symmetries while both $RL$ and $V\!H$ select
the $E_{2g}$ representation. 

Raman scattering was performed in back scattering geometry from the 
$ab$ surface of \MgB2 single crystals grown as described in
\cite{Karpinski}  using less than 2\,mW of incident power focused to 
a $100 \times 200\, \mu$m spot. 
The data in magnetic field was acquired with a continuous flow 
cryostat inserted into the horizontal bore of a SC magnet. 
The sample temperatures quoted have been corrected for laser heating. 
We used the excitation lines of a Kr$^{+}$ laser and a
triple-grating spectrometer for analysis of the scattered light.
The data were corrected for the spectral response of the
spectrometer and the CCD detector and for the optical properties of 
the material at different wavelengths as described in Ref.
\cite{Blumberg94}.

\subsection{Raman response}

\begin{figure}[t]
\includegraphics[width=1\columnwidth]{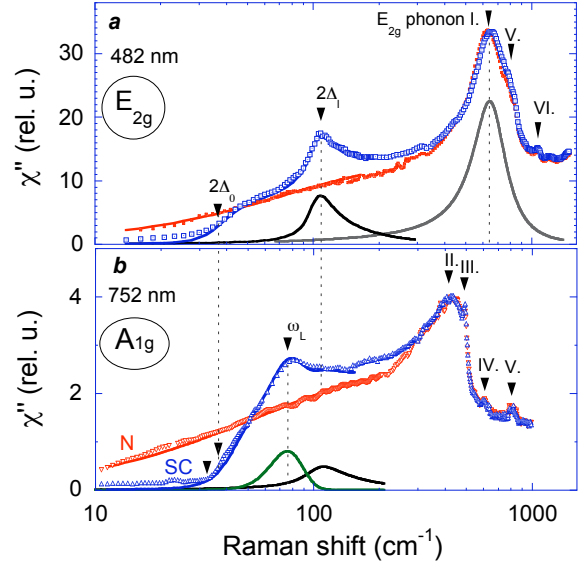}
\caption{
The Raman response spectra of an \MgB2 crystal in 
the normal (red) and SC (blue) states for the \E2g (top) and 
$A_{1g}$ (bottom) scattering channels. 
The data is acquired at 8\,K. 
The normal state has been achieved by applying a 5\,T
magnetic field parallel to the $c$-axis. 
Decomposition into SC coherence peaks, \E2g phonon and the fits are 
shown by solid lines. 
} 
\label{fig:1}
\end{figure}
In Fig.\,\ref{fig:1} we show the Raman response from an \MgB2 
single crystal for the \E2g and \A1g scattering channels
in the normal and SC states.  
The $E_{2g}$ scattering channel is accessed by $RL$ scattering 
polarization geometries and the $A_{1g}$ 
channel by $RR$ geometry. 

The response comprises electronic and phononic contributions. 
The electronic Raman response at low frequencies in the SC state is 
decomposed into a sum (solid lines) of a 
gapped normal state continuum with temperature 
broadened $2\Delta_{0}= 37$\,\cm-1 gap cutoff (threshold at 33\,\cm-1),
the SC coherence peak at $2\Delta_{l} = 109$\,\cm-1 (black solid line), 
and a novel collective mode at $\omega_{L} = 76$\,\cm-1 (green solid line). 
The latter is present only in the $A_{1g}$ scattering channel. 
To fit the observed shapes the theoretical coherence peak singularity 
$\chi^{\prime\prime} \sim 4 \Delta_{l}^{2}/(\omega \sqrt{\omega^{2} - 
4 \Delta_{l}^{2}})$ is broadened by convolution with a Lorentzian with 
HWHM = 12\% of $2\Delta_{l}$ for the \E2g channel and 20\% for 
the \A1g channel. 
The collective mode $\omega_{L}$ is broadened to HWHM = 18\,\cm-1. 

For the high energy part of the spectra the broad \E2g band \emph{I} 
centered at 
about $630-640$\,\cm-1 corresponds to the boron stretching mode which is 
the only Raman active phonon for the \MgB2 structure.  
It is also the only phononic mode demonstrating renormalization 
below the SC transition \cite{Mialitsin}. 
All the other high frequency modes (\emph{II-VI}) in the \A1g and 
\E2g channels correspond to twice the
energy of distinctive flat portions in the phonon dispersions 
measured by inelastic x-ray scattering \cite{Shukla,Baron} and we 
assign them to two phonon scattering. 
The `E$_{1u}$-branch' and a two-fold degenerate low energy acoustic
phonon branch have a coinciding minimum in the A-point of the BZ
thus delivering a large Raman response for the two-phonon peak \emph{II}. 
Peak \emph{III} is due to flat portions of low energy acoustic phonon 
branches when they approach the M-point. 
Peak \emph{IV} is at twice the frequency of a distinctive
saddle point of a high energy acoustic phonon branch in A-point. 
The `A$_{2u}$ branch' is mostly flat all the way along the $\Gamma$-A
line at around $400$\,\cm-1.
This might explain the peculiar
symmetry indifferent behavior of peak \emph{V}. 
Finally the \E2g optical branch has a minimum in the A-point at about 
530\,\cm-1 resulting in the two-phonon scattering peak \emph{VI}.

\subsection{Resonant Raman excitation profile}
\begin{figure}[t]
\includegraphics[width=1\columnwidth]{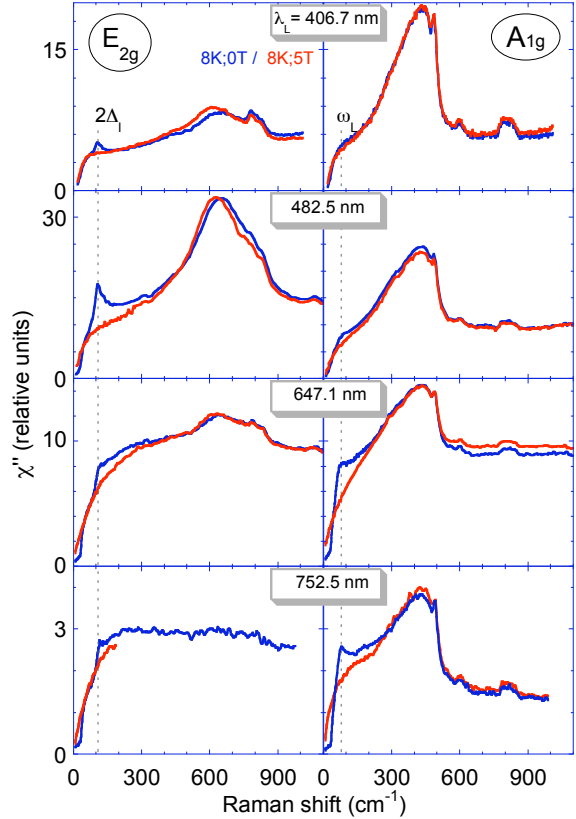}
\caption{
Raman response function at 8\,K in the SC (blue) and 
normal (red) states for the \E2g ($RL$)
and the \A1g ($RR$ polarization) scattering channels as a function of
excitation wavelengths. 
} 
\label{fig:2}
\end{figure}
\begin{figure}[t]
\includegraphics[width=0.95\columnwidth]{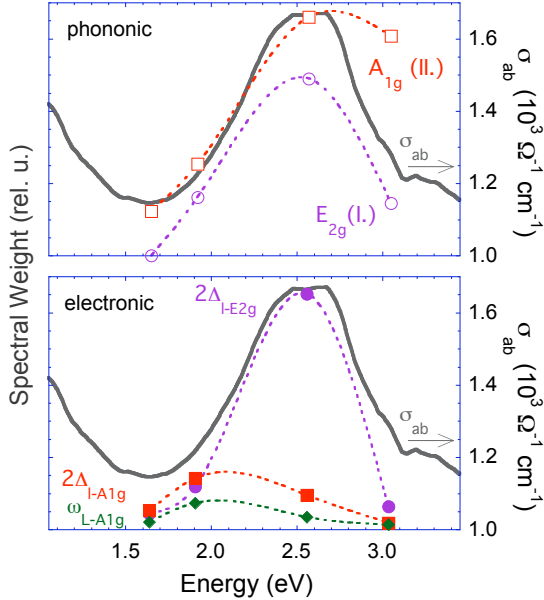}
\caption{
Comparison of $ab$-plane optical conductivity 
(Ref.\cite{Guritanu}) and resonant Raman excitation profiles for 
phononic and electronic excitations. 
The empty symbols show the 340~\cm-1 $A_{1g}$ and the 640~\cm-1 $E_{2g}$ 
phonon intensities and the solid symbols show the SC coherence peaks 
intensities. 
All dashed lines are guides for the eyes. 
} 
\label{fig:3}
\end{figure}

Light can couple to electronic and phononic excitations \emph{via} 
resonant or non-resonant Raman processes \cite{DevereauxRMP}. 
The Raman scattering cross-section can be substantially 
enhanced when the incident 
photon energy is tuned into resonance with optical interband 
transitions. 
The resonance Raman excitation profile (RREP) provides information 
about the scattering probabilities seen in the Raman spectra. 
For \MgB2 the interband contribution to the in-plane optical conductivity  
$\sigma_{ab}(\omega)$ contains strong IR peaks with a tail 
extending to the red part of the visible range and 
a pronounced band around 2.6~eV \cite{Kuz'menko,Guritanu} 
(Fig.\,\ref{fig:3}).   
The IR peaks are associated with transitions between two 
$\sigma$-bands while the 2.6\,eV peak is associated with the $\pi
\rightarrow \sigma$ electronic transitions in the vicinity of the
$\Gamma$ point and $\sigma \rightarrow \pi$ transitions in the
vicinity of the $M$ point of the BZ
\cite{Mazin,Antropov,Kortus}.

To explore the resonance conditions we analyze Raman spectra as a 
function of excitation energy. 
In Fig.\,\ref{fig:2} we show Raman spectra in the SC and normal 
states for the $E_{2g}$ and 
$A_{1g}$ scattering channels for four excitation energies. 
The normal state has been achieved by applying a 5\,T
magnetic field parallel to the $c$-axis.  
All spectra show a relatively strong electronic Raman continuum that 
extends beyond our measurement range. 
The electronic scattering intensity in the fully symmetric $A_{1g}$ 
channel is not much weaker than in the $E_{2g}$ channel indicating 
cancellation of screening that 
could be due to multi-band contributions with opposite sign of the 
effective mass near the FS \cite{Cardona}.  

\section{ELECTRONIC RAMAN RESPONSE}

The low frequency part of the 
electronic Raman continuum changes in the SC state 
(Figs.\,\ref{fig:1}-\ref{fig:2}), reflecting renormalization of electronic 
excitations resulting in three new features in the spectra:  
(i) a threshold of Raman intensity at $2\Delta_{0}= 37$\,\cm-1, 
(ii) a SC coherence peak at $2\Delta_{l} = 109$\,\cm-1, and 
(iii) a new mode at 76\,\cm-1, which is in-between the $2\Delta_{0}$ and 
$2\Delta_{l}$ energies. 
The observed energy scales of the fundamental gap $\Delta_{0}$ 
and the large gap  $\Delta_{l}$ are consistent with 
$\Delta_{\pi}$ and $\Delta_{\sigma}$ as assigned by one-electron 
spectroscopies \cite{Tsuda,Souma,Esklidsen}. 
The features (i-ii) are seen in all scattering geometries while 
mode (iii) contributes only to the $A_{1g}$ scattering channel.  

The Raman coupling to the $2\Delta_{l}$ electronic coherence peak in 
the SC state is provided by density fluctuations in 
the $\sigma$-band.    
For the \E2g channel the peak intensity is enhanced by about an order 
of magnitude when  
the excitation photon energy is in resonance with the 2.6~eV  
$\sigma \rightarrow \pi$ inter-band transitions (Fig.\,\ref{fig:3}).  
In contrast, for the fully symmetric \A1g channel the integrated 
intensity of the $2\Delta_{l}$ coherence peak does not follow the optical 
conductivity and is about five times weaker than for the resonant 
excitation in the \E2g channel. 
Nonetheless, due to relative charge density fluctuations between two 
coupled $\sigma$- and $\pi$-bands the intensity in the fully 
symmetric channel is only partially screened. 
The integrated intensity  of the $\omega_{L}$ collective mode in the 
\A1g channel shows excitation dependence similar to one 
for the $2\Delta_{l}$ coherence peak in the same channel.  

\subsection{The fundamental gap}
At the fundamental gap value $2\Delta_{0}$ the spectra for all  
symmetry channels show a threshold without a coherence peak. 
This threshold appears cleanest for the spectra with lower energy 
excitations for which the low-frequency contribution of 
multi-phonon scattering from acoustic branches is suppressed (see 
Fig.\,\ref{fig:2}). 
The absence of the coherence peak above the threshold is consistent 
with the expected  
behavior for a dirty superconductor \cite{DevereauxRMP}.  
Thus the $\pi$-bands show signatures of strong intrinsic 
scattering leading to the observed Raman continuum.  

The ratio $2\Delta_{0}/k_{B}T_{c}$ is only 1.2 which makes the 
$\pi$-band contribution to the two band superconductivity quite 
tenuous.  
That is in agreement with rapid suppression of the threshold
by a relatively weak magnetic field applied along the $c$-axis. 

\subsection{Large gap in $\sigma$ bands}
The $2\Delta_{l}$ coherence peak is seen for all scattering 
geometries. 
For the \E2g channel it appears as a sharp singularity with  
continuum renormalization extending to high energies, which is in 
agreement with expected behavior for clean 
superconductors \cite{DevereauxRMP}. 
The $2\Delta_{l}$ coherence peak frequency shows a BCS-like temperature
dependence with the $2\Delta^{E}/k_{B}T_{c}$ ratio of about 4
indicating a moderately strong coupling limit. 

Coulomb screening suppresses the scattering intensity 
for the fully symmetric \A1g channel.
The $2\Delta_{l}$ coherence peak intensity does not follow the 
optical conductivity. 
The Raman intensity in the fully 
symmetric channel is governed by the difference in light coupling to 
the the $\pi$- and  $\sigma$-bands which explains the intensity 
enhancement seen for the  pre-resonant excitations (Fig.\,\ref{fig:3}). 
Also, the $2\Delta_{l}$ coherence peak in the \A1g channel 
is broader than in the \E2g channel due to stronger cross 
relaxational coupling to the $\pi$-band quasiparticles.  

\subsection{Leggett's collective mode}
\begin{figure}[t]
\includegraphics[width=1\columnwidth]{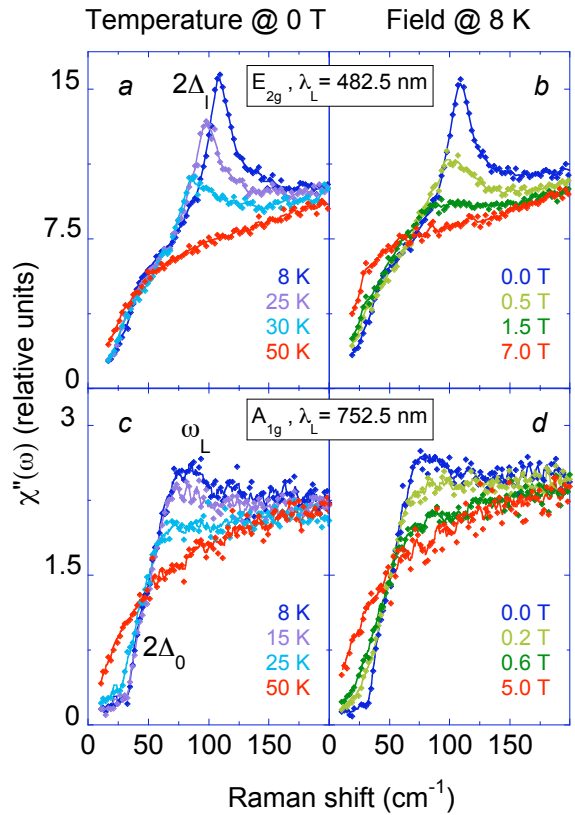}
\caption{
Evolution of low-frequency Raman response as a
function of temperature at zero field and field at 8~K. 
The \E2g scattering channel $(RL)$ with
482.5~nm excitation and the 
\A1g channel $(RR)$ with 752.5~nm excitation are shown. 
} 
\label{fig:4}
\end{figure}
The novel mode at 76\,\cm-1 contributes only to the \A1g 
scattering channel. 
This mode is more pronounced for off-resonance excitation for which 
the electronic continuum above the fundamental threshold 
$2\Delta_{0}$ is weaker. 
We attribute this feature to the collective mode proposed by 
Leggett \cite{Leggett}:
If a system contains two coupled superfluid liquids a simultaneous 
cross-tunneling of a pair of electrons becomes possible. 
Leggett's collective mode is caused by dynamic oscillations of 
Cooper pairs between the two superfluids leading to 
small fluctuations of the relative phase of two condensates while the 
total electron density at every spatial point of the superconductor 
is conserved. 
Such excitation couples to the \A1g Raman scattering channel. 
If the energy of this mode is below the pair-breaking gap, the mode 
dissipation is suppressed and the excitation is expected to be long-lived. 
In the case of \MgB2 the two coupled SC condensates reside at the 
$\sigma$- and $\pi$-bands.

The excitation of Leggett's mode is gapped with a dispersion 
relation for small momentum $q$ given by \cite{Leggett,Griffin,Sharapov} 
\begin{equation}
    \Omega_{L}(q)^{2} = \omega_{L}^{2} + v^{2}q^{2}, 
    \label{dispersion}
\end{equation}
where in the low frequency limit the excitation gap can be expressed 
\emph{via} intra- and 
inter-band pairing potentials $V_{\sigma\sigma}$, $V_{\pi\pi}$ and 
$V_{\sigma\pi}$, 
the gaps $\Delta_{\sigma}$ and $\Delta_{\pi}$ and the density of 
states $N_{\sigma}$ and $N_{\pi}$ in corresponding bands 
\begin{equation}
    \omega_{L}^{2} = \frac{N_{\sigma} + N_{\pi}}{N_{\sigma} 
    N_{\pi}} \frac{4 V_{\sigma\pi} \Delta_{\sigma} 
    \Delta_{\pi}}{V_{\sigma\sigma} V_{\pi\pi} - V_{\sigma\pi}^{2}}. 
    \label{LeggettMode}
\end{equation}
Leggett's mode exists only if 
$V_{\sigma\sigma} V_{\pi\pi} > V_{\sigma\pi}^{2}$. 
The estimates of the coupling constants by first principal 
computations \cite{Liu,Mazin,Choi} show that for the \MgB2 
superconductor this condition is satisfied and the estimate for the 
mode energy is in between 60 - 85\,\cm-1 which is consistent with the 
observed mode at 76\,\cm-1. 
Because the collective mode energy is in between the two-particle 
excitation thresholds 
for the $\pi$- and $\sigma$-bands, $2\Delta_{\pi} < \omega_{L} < 
2\Delta_{\sigma}$, Leggett's excitation rapidly relaxes into 
$\pi$-band quasiparticles. 
Indeed, the measured $Q$-factor for this mode is about two: 
the Cooper pair tunneling energy relaxes 
into $\pi$-band quasiparticle continuum within a couple of tunneling 
oscillations.   
Despite being short lived, Leggett's mode in \MgB2 couples 
to light and is observed by Raman spectroscopy. 

\subsection{Effects of temperature and field}
\begin{figure}[t]
\includegraphics[width=0.98\columnwidth]{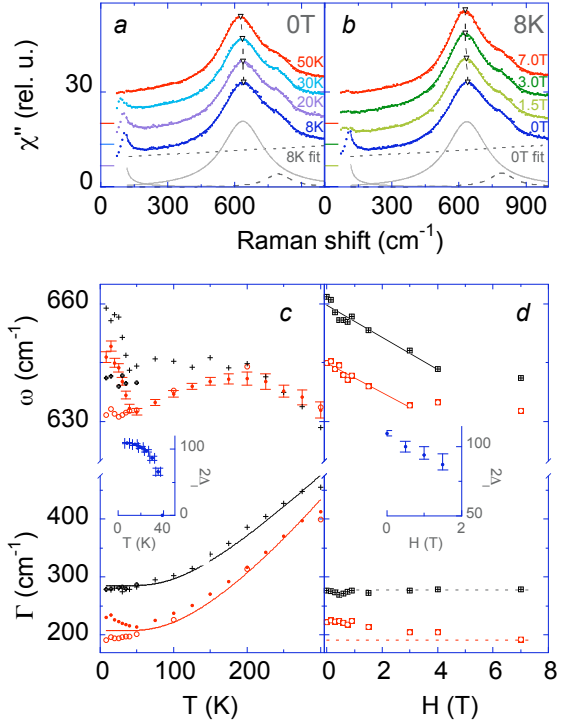}
\caption{
Evolution of the \E2g phonon  with temperature (a) and field (b) 
using 482.5\,nm excitation and ($RL$) polarization. 
The phonon frequency $\omega(T, H)$ and the damping constant 
$\Gamma(T, H)$ are drawn as functions of temperature (c) and field 
(d) for two crystal ${\mathcal A}$ (black) and ${\mathcal B}$ (red 
symbols). 
Insets show temperature and field 
dependencies of the $2\Delta_{l}$ energy.
} 
\label{fig:5}
\end{figure}

In Fig.\,\ref{fig:4} the evolution of the $2\Delta_{l}$ coherence 
peak and Leggett's collective mode $\omega_{L}$
across the SC transition is
displayed for two cases: varying temperature at zero magnetic
field (\textit{a},\,\textit{c}) and varying magnetic field at
8~K (\textit{b},\,\textit{d}). 
The coherence peaks lose their intensity and move to lower energies 
by either increasing temperature or field. 
The intensity threshold $2\Delta_{0}$ is
already smeared out at magnetic fields as weak as 0.2\,T,
consistent with $H^{\pi}_{c2}$ deducted from vortex imaging
\cite{Esklidsen}. 
Leggett's collective mode $\omega_{L}$ persists up
to 0.6\,T while the SC coherence peak $2\Delta_{l}$ is suppressed beyond 2\,T.
$2\Delta_{l}(T, H)$ is shown in the insets of
Fig.\,\ref{fig:5}. 
It exhibits a BCS-like temperature dependence and a linear reduction 
in field with a rapid slope of about -15~\cmT-1.
A linear extrapolation for the $2\Delta_{l}$ gap collapse leads to 
7\,T, a field that is higher than $H^{opt}_{c2}$ \cite{Perucchi}, 
while the coherence peak intensity survives only up to 2\,T.

\section{PHONONIC RAMAN RESPONSE}

High-$T_c$ superconductivity in \MgB2 is known to be promoted
mainly due to the boron layers \cite{Kortus},
thus the high frequency
lattice vibrations of light boron atoms beneficially increase the
electron-phonon coupling.
The \E2g Raman active in-plane boron vibrational mode contributes
significantly to
superconductivity; this fact is reflected by the Eliashberg
function $\alpha^2\,F(\omega)$ peaking in the
same frequency range where a high phononic density of states is
accounted for by Van Hove singularities of the \E2g branch in
the $\Gamma$ and $A$ points of the BZ 
\cite{Yildirim,DagheroPhC}.
The reason the \E2g mode plays a prominent role in the SC 
mechanism is that the mode strongly couples to the
$\sigma$-type states of the boron plane as can be seen from the
basic geometry of the electronic configuration~\cite{Choi}.

\begin{table*}[t]
\caption{
Comparison of $T_{c}$ and the \E2g oscillator parameters for
crystals ${\mathcal A}$ and ${\mathcal B}$.}
\begin{tabular}{cccccccc}
\hline
   & $T_{c}$
   & $\omega_0^{N}$    & $\omega_0^{SC}$ & $\Gamma_{0}$ &
   $\Gamma_{3}$ & $\Gamma_{4}$ & $\kappa$ \\
   Crystal & (K) & (\cm-1) &  (\cm-1) & (\cm-1) & (\cm-1) & (\cm-1) & (\%) \\
\hline\hline ${\mathcal A}$ & 308 & 640 & 659 &
$32\!\pm\!12$ &
$253\!\pm\!10$ & small & $23\!\pm\!4$\\
${\mathcal B}$ & 311 & 630 & 649 & small &
$185\!\pm\!6$ &
$23\!\pm\!3$ & $19\!\pm\!2$\\
\hline
\end{tabular}
\label{tab:AnhDecay}
\end{table*}
Raman spectra exhibit an unusually broad linewidth of the \E2g boron 
stretching mode \cite{Goncharov,Quilty,Mialitsin,RenkerJLTP} which 
has been the subject of numerous speculations. 
While high impurity scattering in earlier low
quality samples has been suggested as one of the possible reasons,
this mechanism can be readily excluded with recent high quality
single crystals. 
The two remaining contributions to the \E2g phonon rapid decay are 
(i) strong electron-phonon coupling and
(ii) multiphononic decay (subsequently referred to as
\emph{anharmonicity}).
The relative importance of the electron-phonon coupling and 
anharmonicity in this matter is still under debate. 
On one hand a density functional theory 
calculation asserts that the anharmonic contribution to
the \E2g phonon linewidth is negligible ($\sim 10$\,\cm-1) \cite{Shukla}.
On the other hand analysis of the phonon self-energy in the long
wavelength limit shows that the $\sigma$-band contribution to the
phonon decay is vanishing \cite{Calandra}.
Thus, even when
contributions of the spectral weight of
$\alpha^2\,F(\omega)|_{\omega < \omega_{E_{2g}}}$ 
to the damping of the \E2g phonon are accounted for \cite{Cappelluti}, 
the experimentally observed linewidth of
$200-280$\,\cm-1 at low temperatures \cite{Quilty,Mialitsin,RenkerJLTP} cannot 
be explained with electron-phonon coupling alone whose part in
the \E2g mode linewidth at low temperatures amounts to about
50\,\cm-1 even in such an elaborate scenario as that in Ref.
\cite{Cappelluti}.

Raman scattering experiments have shown that the frequency of the \E2g mode 
in single crystals at room temperature is around 
635\,\cm-1 \cite{Quilty,RenkerJLTP,Mialitsin} whereas theoretical 
calculations systematically underestimate this value by about
80\,\cm-1 \cite{Yildirim,Shukla}. 
It has been suggested that if the
\E2g band around the $\Gamma$-point is anharmonic then the \E2g mode
frequency is increased by the missing amount to match the
experimentally observed value \cite{Liu,Boeri}. 
In addition, the experimentally observed $T_c$ and the reduced isotope 
effect \cite{HinksPhC} can only be reconciled within anisotropic 
strong coupling theory if the \E2g mode anharmonicity is explicitly 
included \cite{Choi,ChoiPhysRevB}.  

\subsection{Excitation dependence}
The Raman intensities for phononic modes are in resonance with 
the 2.6~eV optical transitions. 
The resonance is more distinct for the \E2g phonon mode that 
reduces by an order of magnitude for adjacent violet and red 
excitations and almost vanishes in the infra-red 
(see Fig.\,\ref{fig:3}) inferring that the Raman coupling to this 
phonon is realized only via $\pi \leftrightarrow \sigma$ interband 
transitions. 
In contrast, the two-phonon scattering in the \A1g channel remains 
visible even for pre-resonance excitations. 

\subsection{Dependence on temperature and field}
In Fig.\,\ref{fig:5}\,(\textit{a-b}) we show the temperature 
dependence of the \E2g Raman response measured on cooling in zero 
field and as a function of field at 8\,K. 
The data (dots) are fitted with two phononic oscillators and a SC 
coherence peak (solid lines) on an electronic continuum 
(decompositions for the lowest spectra are shown). 
In Fig.\,\ref{fig:5}\,(\textit{c-d}) we evaluate the temperature and 
field dependencies of the \E2g phonon frequency $\omega(T, H)$ and 
the damping constant $\Gamma(T, H)$ for two crystals ${\mathcal A}$ 
and ${\mathcal B}$ where we distinguish between the respective values 
for the SC and normal states measured at zero field cooling (solid 
symbols) and 8\,T cooling (empty symbols). 
The solid line in Fig.\,3\,(\emph{c}) is a fit of the
damping constant $\Gamma(T)$ in the normal state to a model of
anharmonic two and three phonon decay at one-half and one-third
frequencies:
\begin{eqnarray}
    \Gamma(T) = & \Gamma_{0} +
\Gamma_{3}[1 + 2n(\Omega(T)/2)] + \nonumber \\
 & \Gamma_{4}[1 +
3n(\Omega(T)/3)+3n^2(\Omega(T)/3)].
\label{damping}
\end{eqnarray}
Here $\Omega(T)={h\,c\, \omega_{h}}/{k_{B}T}$, with the harmonic
frequency $\omega_{h} = 540$\,\cm-1 
\cite{Mazin,Shukla,Kortus}, $n(x)$ is
the Bose-Einstein distribution function, $\Gamma_{0}$ is the
internal temperature independent linewidth of the phonon, and
$\Gamma_{3,4}$ are broadening coefficients due to the cubic and
quartic anharmonicity. The results of the fit to this anharmonic
decay model are collected in Table \ref{tab:AnhDecay}. 
For both
crystals the broadening coefficients $\Gamma_{3} + \Gamma_{4} \gg
\Gamma_{0}$ and therefore the anharmonic decay is primarily
responsible for the large damping constant of the \E2g phonon. 
We identify the reason for this rapid phononic decay in the phononic
density of states (PDOS) peaking at 265\,\cm-1, half of the harmonic
\E2g phonon frequency $\omega_h$ (Refs. \cite{Yildirim,Osborn}), 
which corresponds to the Van-Hove 
singularity of the lower acoustic branch, 
almost dispersionless along the $\Gamma-{\rm K}-{\rm M}$ direction. 
In this context the narrowing of the \E2g mode
with Al substitution observed in Refs.~\cite{RenkerJLTP,Bohnen} can 
be readily explained with the \E2g phonon branch moving to energies 
above 100\,meV with increased Al concentration
whereas the acoustic modes that provide the decay channels stay close
to their original energies with high PDOS in the energy range of 
$200-320$\,\cm-1 \cite{Bohnen}.
In short, the fast decay of the \E2g mode
is due to the unique combination of its harmonic frequency
in the $\Gamma$ point corresponding to high PDOS at half of 
this frequency. 
The residual
linewidth $\Gamma_0$ that we obtain from the fit to the
anharmonic decay model, while small, is not in contradiction with the 
theoretical estimates \cite{Calandra,Cappelluti} of the
electron-phonon decay contribution to the \E2g phonon linewidth.

It is worth noting that individual $\Gamma_i$ parameters differ for 
the two single crystals despite the fact that both
samples were grown in the same batch.
The \E2g mode for crystal ${\mathcal A}$ is broader by about 10\,meV 
than for crystal ${\mathcal B}$.
With $\Gamma_0^{\mathcal A}$
somewhat higher than $\Gamma_0^{\mathcal B}$ and
$\Gamma_3^{\mathcal A}$ substantially higher than
$\Gamma_3^{\mathcal B}$ (see Table\,\ref{tab:AnhDecay}) the \E2g
mode in crystal ${\mathcal A}$ is more anharmonic than in crystal
${\mathcal B}$. Accordingly the crystal ${\mathcal A}$ mode is
pushed to about 10\,\cm-1 higher frequency at low temperatures. We
note a correlation between the larger anharmonicity and slightly lower
$T_{c}$\,in the case of crystal ${\mathcal A}$.

\subsection{Pressure and Al substitution}
\begin{figure}[t]
\begin{center}
\includegraphics[width=0.64\columnwidth]{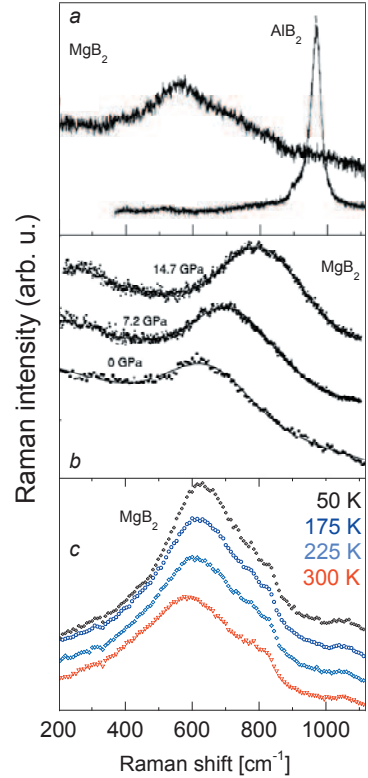}
\end{center}
\vspace{-30pt}
\caption{
The phononic Raman intensity in \E2g channel as function of 
substitution, pressure and temperature. 
(a) \MgB2 vs. AlB$_{2}$ at room temperature from Ref.\,\cite{Bohnen}; 
(b) The pressure dependence from Ref.\,\cite{Goncharov};
(c) The temperature dependence at ambient pressure. 
Spectra are shifted vertically for clarity.
} 
\label{fig:6}
\end{figure}
The boron stretching \E2g phonon has been found to be the most 
sensitive mode to structural changes upon substitution of Mg sites with Al. 
The Raman spectra of gradual substitution Al$_{x}$Mg$_{1-x}$B$_{2}$ 
are quite complicated with nonuniform transfer of spectral weight from the
640\,\cm-1 mode as observed in pure \MgB2 to the 980\,\cm-1 AlB$_{2}$ 
\E2g mode \cite{RenkerJLTP}. 
Upon complete substitution the change in shape and frequency is 
striking as the \E2g mode in AlB$_{2}$ has stiffened
more than 300\,\cm-1 and its line width has narrowed from 400\,\cm-1
to about 50\,\cm-1 (see Fig 6\,(\emph{a})). 
With Al substitution the large anharmonicity of the
\E2g phonon mode is reduced.

The pressure dependence of the \E2g phonon frequency links its 
frequency shift to the variation of the lattice parameters 
\cite{Goncharov}. 
In the pressure range up to 15\,GPa the mode frequency shifts linearly with
pressure (Fig.\,6\,(\emph{b})). 
An unusually large Gr\"ueneisen parameter of 3.9 has been deduced from 
this frequency shift \cite{GoncharovPhysicaC}.
The larger Gr\"ueneisen parameters are usually related to increased 
anharmonicity of the mode \cite{Zallen} which fits in the overall 
picture of anharmonicity as discussed above.
Also interesting is the link to the linear decrease of $T_{c}$ with 
pressure which in the above mentioned pressure range 
suppresses $T_{c}$  \cite{Tomita,Loa}

The temperature dependence between room temperature and $T_{c}$ is a
smooth but nonmonotonic function peaking at 200 K (fit results to
the spectra are shown in Fig.5\,(\emph{c})). 
Its particular functional shape reflects the variation of 
anharmonicity of the \E2g as a function temperature as both the \E2g 
band and the corresponding decay channels react to lattice expansion 
with increased temperature.

\subsection{Phononic self-energy effects}
To describe the superconductivity induced self-energy effect we
refer to Fig.\,\ref{fig:5}\,(\emph{c}).
Upon cooling in zero field the \E2g phonon frequency exhibits 
nonmonotonic but smooth behavior down to $T_c$. 
Then at $T_c$ it displays
abrupt hardening with $\omega_0^{SC}(T)$ scaling to the functional
form of the SC gap magnitude $2\Delta_{l}(T)$. 
For in-field cooling the \E2g phonon frequency $\omega_0^N(T)$ remains
unrenormalized. 
The differences between the phonon frequencies in
the normal and SC states at 8\,K are
$18 \pm 3$ and $15 \pm 1.6$\,\cm-1 for crystals  ${\mathcal A}$
and ${\mathcal B}$ respectively.
To quantify the relative hardening of the \E2g mode
we obtain the superconductivity induced renormalization
constant $\kappa = (\omega_0^{SC}/{\omega_0^{N}}) - 1 \approx 2.5\%$ 
(see Table\,\ref{tab:AnhDecay}) which 
is much smaller than the theoretically predicted
$\kappa \approx 12$\% \cite{Liu}.

We estimate the electron-phonon coupling constant
$\lambda^{\Gamma}_{E_{2g}}$ around the BZ center using
approximations adopted in Refs.~\cite{Zeyher,Rodriguez}: $\lambda
= -\kappa \, {\mathcal Re}\,(\frac{\sin u}{u})$, where $u \equiv
\pi + 2 i \cosh^{-1}({\omega^{N}}/{2\Delta_{\sigma}})$, and obtain
$\lambda^{\Gamma}_{E_{2g}} \approx 0.3$.
This estimate of the coupling constant is consistent with the fit to 
a phenomenological model \cite{Zeyher2} where the direct coupling of 
light to the $\sigma$ bands is neglected but it is  
smaller than the values predicted by the first principal 
computations \cite{Liu,Mazin,Kortus,Choi,Golubov}. 

We note that all the other modes contributing to two-phonon 
scattering do not exhibit any measurable renormalization upon 
cooling into the SC state (see Figs.\,\ref{fig:1}-\ref{fig:2}), 
thereby the 635\,\cm-1 \E2g boron stretching 
mode is the only phonon that exhibits renormalization below the SC 
transition. 

\section*{SUMMARY}
We have measured the polarization resolved Raman response as a function of 
temperature, field and excitation energy for \MgB2 single crystals. 

The electronic scattering data revealed three superconductivity 
induced spectroscopic features: 
a clean threshold below $2\Delta_{0}=37$\,\cm-1 corresponding to the 
fundamental gap,  
a coherence peak at $2\Delta_{l}=109$\,\cm-1 corresponding to the gap 
on the $\sigma$-bands FS, 
and the Leggett's collective mode at $\omega_{L}=78$\,\cm-1 arising from the 
fluctuation in the relative phase between two coupled SC 
condensates residing on two bands. 
Altogether the electronic Raman spectra show signatures for 
superconductivity in the clean limit for quasi-two-dimensional 
$\sigma$-bands and dirty limit for three-dimensional 
$\pi$-bands.  
The ratio $2\Delta_{0}/k_{B}T_{c}$ is only 1.2 which makes the 
$\pi$-band contribution to the two band superconductivity quite 
tenuous, in agreement with rapid suppression of the threshold 
frequency by a relatively weak magnetic field. 
The large gap shows a BCS-like temperature
dependence with the $2\Delta_{l}/k_{B}T_{c}$ ratio of about 4
indicating a moderately strong coupling limit. 
The $2\Delta_{l}$
gap magnitude is suppressed by an external magnetic field at the
rapid rate of -15~\cmT-1. 

From the
temperature dependence of the \E2g boron stretching phonon we
conclude that anharmonic decay is primarily responsible for the
anomalously large damping constant of this mode.
For this phonon we observe a SC induced self-energy effect and
estimate the electron-phonon coupling constant.

\subsection*{Acknowledgments}
The authors thank M.\,V.~Klein, A.\,A. Kuz'menko, D.\,van\,der\,Marel, 
I.\,I.~Mazin and W.\,E.~Pickett for valuable discussions. 
AM was supported by the Lucent-Rutgers Fellowship program.
NDZ was supported by the Swiss National Science Foundation through 
NCCR pool MaNEP.

\end{document}